\documentstyle[11pt]{article} 

\newtheorem{th}{Theorem}  
\newtheorem{ax}{Axiom}  
\newtheorem{lm}{Lemma} 
\newtheorem{df}{Definition}      
\newtheorem{pr}{Proposition}      
\newtheorem{cl}{Conclusion}  
\newtheorem{re}{Remark}    
\newtheorem{as}{Assumption}  
\newtheorem{wg}{Wild Guess}
\newtheorem{ex}{Example}
\newcommand{\bth}{\begin{th}\hspace{-5pt}{\bf .} \ }    
\newcommand{\eth}{\end{th}}
\newcommand{\bax}{\begin{ax}\hspace{-5pt}{\bf .} \ } 
\newcommand{\eax}{\end{ax}}
\newcommand{\blm}{\begin{lm}\hspace{-5pt}{\bf .} \ }  
\newcommand{\elm}{\end{lm}}
\newcommand{\bdf}{\begin{df}\hspace{-5pt}{\bf .} \ }      
\newcommand{\edf}{\end{df}} 
\newcommand{\bpr}{\begin{pr}\hspace{-5pt}{\bf .} \ }     
\newcommand{\epr}{\end{pr}}
\newcommand{\bcl}{\begin{cl}\hspace{-5pt}{\bf .} \ }   
\newcommand{\ecl}{\end{cl}}
\newcommand{\bre}{\begin{re}\hspace{-5pt}{\bf .} \ }
\newcommand{\ere}{\end{re}}
\newcommand{\bas}{\begin{as}\hspace{-5pt}{\bf .} \ }
\newcommand{\eas}{\end{as}}
\newcommand{\bwg}{\begin{wg}\hspace{-5pt}{\bf .} \ }
\newcommand{\ewg}{\end{wg}}  
\newcommand{\bex}{\begin{ex}\hspace{-5pt}{\bf .} \ }
\newcommand{\eex}{\end{ex}}
\newcommand{\bpf}{\noindent {\it Proof:} }
\newcommand{\epf}{\hfill$\bullet$\par\vspace{3mm}\noindent}  
\newcommand{\bit}{\begin{itemize}}
\newcommand{\eit}{\end{itemize}\par\noindent}
\newcommand{\beq}{\begin{equation}} 
\newcommand{\eeq}{\end{equation}\par\noindent}
\newcommand{\beqa}{\begin{eqnarray*}}
\newcommand{\eeqa}{\end{eqnarray*}\par\noindent} 
\newcommand{\beqn}{\begin{eqnarray}}  
\newcommand{\eeqn}{\end{eqnarray}\par\noindent}
  
\begin{document}     
\parindent=1.0cm  
{\bf STRUCTURAL CHARACTERIZATION OF COMPOUNDNESS\footnote{Appeared in {\it Int. J. Theor. Phys.} {\bf 39},
585--594 (2000), as the proceedings of a lecture at the fourth biannual conference of the International
Quantum Structures Association, Liptovsk\'y J\'an, Slovakia, 1998\,; a more elaborated and detailed
version of this paper is in preparation.}}      
\par
\vspace{0.3cm}   
\par {Bob Coecke\footnote{The author thanks David Moore, Isar Stubbe and Frank Valckenborgh
for discussion on the subject of this paper and Constantin Piron for his hospitality at
Universit\'e de Gen\`eve.}}
\par\vskip 0.3 truecm\par  
{\it\scriptsize
FUND-DWIS,  Free University of Brussels,
Pleinlaan 2,  B-1050 Brussels, Belgium; bocoecke@vub.ac.be}
\par\vskip 0.3 truecm\par  
{\it Abstract:} 
We recover the rays in the tensor product of Hilbert spaces
within a larger class of so called `states of compoundness', 
structured as a complete lattice with the `state
of separation' as its top element. At the base of the construction lies the assumption that the
cause of actuality of a property of one of the (as individual considered) entities in the compound system
can be actuality of a property of the other one.
\parindent=0.8cm
\par\bigskip\par
\noindent 
{\bf 1. INTRODUCTION}       
\par\bigskip\par  
\noindent 
Most approaches towards a realistic description of compound ---quantum--- systems are based on the
recognition of subsystems, imposing some mathematical universal property as a structural
criterion (Hellwig and Krauser, 1977; Zecca, 1977; Aerts and Daubechies, 1978; Aerts, 1982;
Pulmannov\'a, 1984; Ischi, 1999; Valckenborgh, 2000).  In this paper we take a different point of
view, essentially focusing on a structural characterization of the interaction 
between the individual entities, rather than on the compound system as a whole. 
More precisely, we structurize the concept of `mutual induction of actuality'  for the
`individual entities' in a compound system, inspired by the existence of an ---essentially
unique--- representation for compound quantum systems when postulating that a state transition of
one individual entity induces a state transition of the others ---see Coecke (1998a).    In
particular will we consider `separation' as one particular state of compoundness,
and not as a type of `entity' as in Aerts (1982) and Ischi (1999), as such avoiding some
axiomatic drawbacks that emerge when taking the latter perspective. Formally, an essential
ingredient of the reasoning can be borrowed from Faure {\it et al.} (1995) where propagation
of states and properties in maximal deterministic evolutions is studied.  As an
application of our way of looking at compoundness we mention a representation for spin systems
(Coecke 1995, 1998b).  The mathematical preliminaries to this paper are basic notions on linear operators for which we refer
to Weidmann (1981) and that of a Galois
dual pair, i.e., a couple of isotone maps $f:M\to N$ and
$f^*:N\to M$ satisfying $\forall x\in M, y\in N:x\leq f^*(y)\Leftrightarrow
f(x)\leq y$, where $f$ preserves existing joins, $f^*$ preserves existing meets, and
we have existence and uniqueness of it for a meet (resp$.$ join) 
preserving map between complete lattices (Birkhoff,
1940; Johnstone, 1982).
\vfill\eject\noindent 
{\bf 2. COMPOUNDNESS AND ASSIGNMENT OF CAUSES}     
\par\bigskip\par
\noindent 
Let us first recall the general concept of an {\it entity}, along the lines of Jauch and Piron (1969),
Piron (1976) and Aerts (1982) ---we will not go into the details on this and refer
for the most recent overview to Moore (1999). We consider an entity to be a physical system
described by a collection of either potential or actual properties ${\cal L}$, partially ordered by an
---operationally motivatable--- implication relation `$\leq$', and which {\it proves} to be a complete
lattice (Piron, 1976).  As is discussed in Piron (1977) and Moore (1999), the meet `$\wedge$' can be
treated as a classical conjunction. A property is said to be `actual' if we get {\it true}
with certainty when it would  be verified in any possible way; $a\in {\cal L}$ is stronger (resp$.$
weaker) than
$b\in{\cal L}$ iff
$a\leq b$ (resp$.$
$b\leq a$); the top element  
$1$ of the complete lattice can be seen as expressing `existence' of the entity and the
bottom element
$0$ expresses the `absurd', i.e., what can never be true.  As an example, the property lattice ${\cal
L}_{\cal H}$ of a quantum entity described in a Hilbert space ${\cal H}$ is the set of closed subspaces
ordered by inclusion with intersection as meet and closed linear span as join. 
We will also
systematically use the term `individual entity' when considering identifiable `parts' in a larger
system ---as such to be seen as a compound system--- since in general, these individual entities do
not satisfy the general conception of what an entity is in the references  mentioned above ---for a
discussion on this aspect see Coecke (1998a).  In the presence of interaction
between individual entities, actuality of a property
$a_2\in{\cal L}_2$ of individual entity $S_2$ might be due to the actuality of a property
$a_1\in{\cal L}_1$ of individual entity $S_1$.  In particular, we will show that all interaction
involved in quantum entanglement can be expressed in this way, and therefore we define a map:  
\beq
f^*:{\cal L}_2\to{\cal L}_1:a_2\mapsto``the\ cause\ in\ {\cal L}_1\ of\ the\ actuality\ of\ a_2"  
\eeq
This cause of actuality of $a_2$ is the
{\it weakest $a_1\in{\cal L}_1$ that assures actuality of $a_2$}: indeed, any $b_1\in{\cal L}_1$
with  $b_1\leq a_1$ then automatically causes actuality of $a_2$ since it implies $a_1$. Note
that existence of such a weakest
$a_1\in{\cal L}_1$ follows from the fact that `assuring actuality of $a_2$' 
precisely defines it as a property for $S_1$.  As an example, when considering two
separated individual entities it is the bottom element $0_1\in{\cal L}_1$ that assures actuality of
any
$a_2\in{\cal L}_2\setminus\{1_2\}$, explicitly expressing that no state of ${\cal L}_1$ assures
anything about
${\cal L}_2$. On the contrary, the property $1_1$ assures actuality of $1_2$ since we {\it a priori} assume the
existence of both individual entities.  
When considering meets $\wedge_ia_{2,i}$ in ${\cal L}_2$, due to their
significance as a classical conjunction, i.e. they can be read as `and', $\wedge_if^*(a_{2,i})$ assures
actuality of
$\wedge_ia_{2,i}$, as such assuring $f^*$ to preserve non-empty meets. Since
$\wedge\emptyset=1$ and
$f^*(1_2)=1_1$ by assumption of the existence of both individual entities, 
it also preserves the empty meet.  Thus, there
exists a unique join-preserving Galois dual for      
$f^*$, namely:  
\beq\label{Galois}  
f:{\cal L}_1\to{\cal L}_2:a_1\mapsto\wedge\{a_2\in{\cal L}_2|a_1\leq f^*(a_2)\}=min\{a_2\in{\cal L}_2|a_1\leq
f^*(a_2)\}
\eeq
From eq.(\ref{Galois}) it follows that $f$ assigns to a property $a_1\in{\cal L}_1$ the strongest
property
$a_2\in{\cal L}_2$ of which it assures actuality ---the minimum of all $a_2\in{\cal L}_2$ such that
$a_1\leq f^*(a_2)$--- implicitly implying actuality of all
$b_2\geq a_2$. Thus, 
$f$ expresses exactly {\it induction of actuality of ${\cal L}_1$ 
on ${\cal L}_2$}. 
\bcl
A `state of compoundness for $S_1$ on $S_2$' is a join preserving map $f:{\cal    
L}_1\to{\cal L}_2$.  
\ecl
\par\smallskip\par 
\noindent 
{\it i) STRUCTURING STATES OF COMPOUNDNESS:}     
\par\smallskip\par  
\noindent 
Denote by $Q({\cal L}_1,{\cal L}_2)$ the join preserving maps from
${\cal L}_1$ to
${\cal L}_2$ and set for all $\{f_i\}_i\subseteq Q({\cal L}_1,{\cal L}_2)$: 
\beq\label{bigvee}
\bigvee_i f_i:={\cal L}_1\to{\cal L}_2:a\mapsto\vee_if_i(a)
\eeq    
Eq(\ref{bigvee}) defines a complete internal operation on $Q({\cal L}_1,{\cal L}_2)$ since 
for $\{f_i\}_i\subseteq Q({\cal L}_1,{\cal L}_2)$:
$({\bigvee}{}_if_i)(\vee_ja_j)=\vee_{ij}f_i(a_j)=\vee_j(\bigvee_if_i)(a_j)$, and one easily
verifies that
$\bigvee_if_i$ is the least upper bound of $\{f_i\}_i$ in $Q({\cal L}_1,{\cal L}_2)$. 
\bcl
The states of compoundness for $S_1$ to $S_2$ are described by a complete 
lattice $(Q({\cal L}_1,{\cal
L}_2),\bigvee)$, inheriting its join from the underlying property lattice ${\cal L}_2$
pointwisely.  
\ecl 
Analogously, the collection of meet preserving maps $f^*:{\cal L}_2\to{\cal L}_1$ denoted by
$Q^*({\cal L}_1,{\cal L}_2)$ is a meet complete lattice with respect to pointwise meet, denoted as
$\bigwedge$. Note here that the significance of the
lattice meet as a classical conjunction is lifted by the pointwise computed meets to the
level of assignment of temporal causes, as such giving to $f^*\bigwedge g^*$ the significance
of `$f^*$ {\it and} $g^*$'.
Set $L^*(f,a_2)=\{a_1\in {\cal L}_1|f(a_1)\leq a_2\}$ and $L(f^*,a_1)=\{a_2\in {\cal L}_2|a_1\leq
f^*(a_2)\}$. If
$\forall a_1\in {\cal L}_1:f(a_1)\leq g(a_1)$, then $[g(a_1)\leq a_2\Rightarrow f(a_1)\leq a_2]$
and 
$\forall a_2\in {\cal L}_2:L^*(g,a_2)\subseteq L^*(f,a_2)$.
Thus, $\vee L^*(g,a_2)\leq\vee L^*(f,a_2)$
yielding $\forall a_2\in {\cal L}_2:g^*(a_2)\leq f^*(a_2)$ and
$g^*\leq f^*$. Reversely, $\forall a_2\in {\cal L}_2:g^*(a_2)\leq f^*(a_2)$ implies $[g^*(a_2)\geq
a_1\Rightarrow f^*(a_2)\geq a_1]$, $L(g^*,a_1)\subseteq L(f^*,a_1)$, 
$\wedge L(g^*,a_1)\geq\wedge L(f^*,a_1)$, and thus $f\leq g$. As such $f\leq g$ iff $g^*\leq f^*$.
It follows that $(Q^*({\cal L}_1,{\cal L}_2),\bigwedge)^{op}$ and $(Q({\cal
L}_1,{\cal L}_2),\bigvee)$ are isomorphic complete lattices ---`$(-)^{op}$' stands for reversal of
partial order.   
\bcl
The map $\ ^*:Q({\cal L}_1,{\cal L}_2)\to Q^*({\cal L}_1,{\cal L}_2):f\mapsto f^*$ `interprets'
$\bigvee_if_i$, the join of a set of states of compoundness, as $\bigwedge_if_i^*$ a conjunction of
the corresponding assignments of temporal causes.
\ecl
\par\smallskip\par
\noindent   
{\it ii) SEPARATION AS THE TOP STATE OF COMPOUNDNESS:}  
\par\smallskip\par  
\noindent 
Since we have $f^*:\{{\cal L}_2\setminus\{1_1\}\to{\cal L}_1:a_2\mapsto 0_1\, ;\, 1_2\mapsto 1_1\}$ in case of
separation, the `state of separation' for $S_1$ on $S_2$ is given by $f:\{{\cal
L}_1\setminus\{0_2\}\to{\cal L}_2:a_1\mapsto 1_2\, ;\, 0_1\mapsto 0_2\}$ and this is exactly the top
element $1_{1,2}$ of $Q({\cal L}_1,{\cal L}_2)$ ---note that
$0_1\mapsto 0_2$ is required for any join preserving map, the bottom being the empty join. Remark that the
bottom element
$0_{1,2}$ of $Q({\cal L}_1,{\cal L}_2)$ stands for the `absurd state of compoundness'. Indeed, since
$f:{\cal L}_1\to{\cal L}_2:a_1\mapsto 0_1$ we have $f^*:{\cal L}_2\to{\cal L}_1:a_2\mapsto 1_1$, i.e., 
existence of
$S_1$ causes actuality of all properties of $S_2$, and as such actuality of the absurd property $\wedge{\cal
L}_2=0_2$.  
\par\bigskip\par 
\noindent 
{\it iii) QUANTUM ENTANGLEMENT AS ATOMIC STATES OF COMPOUNDNESS:}  
\par\smallskip\par  
\noindent 
We will now consider those $f\in Q({\cal L}_{{\cal H}_1},{\cal L}_{{\cal H}_2})$
that send atoms to atoms or $0_2$ for ${\cal L}_{{\cal H}_1}$ and ${\cal L}_{{\cal H}_2}$ the lattices
of closed subspaces of Hilbert spaces. The following
result can be found in Faure {\it et al.} (1995) and is essentially based on Faure and Fr\"olicher (1993,
1994) and Piron (1964, 1976): {\it Any non-trivial}\,\footnote{I.e., with at least three non-collinear
elements in its range. Note that if the image is spanned by either one or two atoms there is an
obvious  representation as a linear map, extending the collection of representations in a natural way,
which motivates us to assume a linear representation on the underlying Hilbert space for any state of
compoundness.} {\it $f\in Q({\cal L}_{{\cal H}_1},{\cal L}_{{\cal H}_2})$
that sends atoms to atoms or $0_2$
induces either a linear or an anti-linear map $F:{\cal H}_1\to{\cal H}_2$.}
\bpr
If $f\in Q({\cal L}_{{\cal H}_1},{\cal L}_{{\cal H}_2})$ sends atoms to atoms or\, $0_2$ than
$f$ is itself an atom or $0_{1,2}$.    
\epr
\bpf  
Let $K_f$ and $K_g$ be the respective kernels of the linear maps $F,G:{\cal H}_1\to{\cal H}_2$
induced by
$f,g\in Q({\cal L}_{{\cal H}_1},{\cal L}_{{\cal H}_2})$ with $f<g$, and  
thus, $K_g\subset K_f$ and $K_f^\perp\subset K_g^\perp$. For $\psi\in
K_f^\perp\setminus\{\underline{0}_1\}$ and
$\phi\in (K_f\cap K_g^\perp)\setminus\{\underline{0}_1\}$ we have $F(\psi+\phi)=F(\psi)=G(\psi)$ whereas
$G(\psi+\phi)=G(\psi)+G(\phi)$, forcing
$G(\psi)=kG(\phi)$ for $k$ non-zero. However, then $G(k\psi-\phi)=0$ although $k\psi-\phi\in
K_g^\perp$, yielding contradiction except when $K_f^\perp=\underline{0}_1$, i.e., $[f:a_1\mapsto
0_2]=0_{1,2}$.
\epf
For the sake of transparency of the argument we will from now on only consider finite
dimensional Hilbert spaces. Let
${\cal H}'_1$ be the Hilbert space of continuous linear functionals on ${\cal H}_1$, connected to it by the
correspondence ${\cal H}_1\to{\cal H}'_1:\psi\mapsto\langle\psi|-\rangle$.  Then the tensor
product ${\cal H}'_1\otimes{\cal H}_2$ is isomorphic to the space of
linear operators with indicated domain and codomain ---denoted as $B({\cal H}_1,{\cal H}_2)$--- by the
isomorphism: 
\beq
{\cal H}'_1\otimes{\cal H}_2\to B({\cal H}_1,{\cal
H}_2):\Bigl[\sum_{i=1}^mc_i\langle\psi_i|-\rangle\otimes
\phi_i\Bigr]\mapsto\Bigl[F_{\{c_i\}_i}:{\cal H}_1\to{\cal
H}_2:\psi\mapsto\sum_{i=1}^mc_i\langle\psi_i|\psi\rangle\phi_i\Bigr]
\eeq
with $\{\psi_i\}_i$ and $\{\phi_i\}_i$ fixed orthonormal bases
---note    
that
$||F_{\{c_i\}_i}||_{HS}=||\sum_{i=1}^mc_i\langle\psi_i|-\rangle\otimes\phi_i||_{{\cal H}'_1\otimes{\cal
H}_2}$ for
$||-||_{{\cal H}'_1\otimes{\cal H}_2}$ the Hilbert space metric and
$||-||_{HS}$ the Hilbert-Schmidt norm. Considering anti-linear maps
$\sum_{i=1}^mc_i\langle-|\psi_i\rangle\phi_i$ ---say $B'({\cal H}_1,{\cal H}_2)$--- with a
reasoning along the same lines, we obtain
${\cal H}_1\otimes{\cal H}_2$. As such, we recover the rays of the
tensor product of Hilbert spaces
${\cal H}_1\otimes{\cal H}_2$ as a special case of our more general class of atomic states of
compoundness, besides
the rays ${\cal H}'_1\otimes{\cal H}_2$. Indeed, although ${\cal
H}_1\otimes{\cal H}_2$ and
${\cal H}'_1\otimes{\cal H}_2$ are isomorphic as Hilbert spaces, 
there rays represent different
states of compoundness. 
\par\bigskip\par    
\noindent 
{\bf 3. COMPOUNDNESS AS EVOLUTION --- AND VICE VERSA}  
\par\bigskip\par  
\noindent 
Although we were able to use much material from Faure {\it et al.} (1995) ---that deals with evolution
of an entity--- we should note that an essential difference
between the two formal developments is  due to the fact that existence of the entity at a certain instance
of time in general does not assure existence in the future when considering evolution. However, imposing
a condition on the `type' of evolution that we consider by `requiring preservation' makes an
illustrative comparison possible, and as such we will proceed. To fix ideas, identify ${\cal L}_1$
with the property lattice of a fixed entity at time
$t_1$ and ${\cal L}_2$ as its property lattice at time $t_2$. We can again consider a map $f^*:{\cal
L}_2\to{\cal L}_1$ that assigns to a property that is actual at time $t_2$ the cause of its actuality
at time
$t_1$, and the corresponding Galois dual $f:{\cal L}_1\to{\cal L}_2$ that now expresses 
temporal propagation
of properties, all of them again structured in $Q({\cal L}_1,{\cal L}_2)$.   
\par\smallskip\par      
\noindent 
{\it Example 1}: A `maximal deterministic evolution' is defined as one where atoms propagate to
atoms or
$0_2$ ---the latter in order to express a domain for initial states--- mirroring atomic states of
compoundness.  
\par\smallskip\par      
\noindent 
{\it Example 2}:  We could define a `maximal indeterministic ---or minimal
deterministic--- evolution' as one that strictly assures existence at time $t_2$, yielding
the mirror of the state of separation.  
\par\smallskip\par      
\noindent 
For a more elaborated  
formal discussion on the connection between compoundness and evolution we refer to Coecke and
Stubbe (1999a) and Coecke and Moore (2000).  
\par\bigskip\par
\noindent 
{\bf 4. ON ASSIGNMENT OF PROPER STATES TO INDIVIDUALIZED ENTITIES}         
\par\bigskip\par
\noindent 
\par\smallskip\par                
\noindent 
The discussion on evolution was not merely illustrative but constitutes an essential ingredient in this
section, where we discuss state transitions of individual entities due to a state of compoundness: in
Section 2 we characterized entanglement between individual entities in a compound system, but no
consideration on a characterization of individual entities themselves has been made.   Indeed, a
description of a compound system consisting of two individual entities $S_1$ and $S_2$ requires a 
characterization of these individual entities themselves besides the states of
compoundness $f_{1,2}$ of $S_1$ on $S_2$ and
$f_{2,1}$ of $S_2$ on $S_1$.  However, this {\it proves} to requires a more general concept of
state than in Piron (1976), Aerts (1982) and Moore (1995), and a different name {\it proper state} has
been introduced in Coecke (1998a) to stress this difference. Therefore consider for each individual
entity an {\it a priori} set of proper states
$\Sigma$ and denote by
${\cal C}:{\cal P}(\Sigma)\to{\cal L}$ the map that assigns to any $T$ in ${\cal P}(\Sigma)$, the
powerset of
$\Sigma$, the strongest property in ${\cal L}$ that is implied by every $p\in\ T$. As shown in  
Coecke and Stubbe (1999a, 1999b), ${\cal C}$ canonically induces a pre-order on $\Sigma$ by
$p\leq_{\cal C}q\Leftrightarrow{\cal C}(\{p\})\leq{\cal C}(\{q\})$. Moreover, the above discussed
requirement that properties propagate with preservation of
join (see also Pool, 1968) restricts the ---not-necessarily deterministic--- state
transitions to be described by a map in:  
\beq
Q^\#(\Sigma)=\bigl\{f:{\cal P}(\Sigma)\to{\cal P}(\Sigma)\bigm|f(\cup T)=\cup f(T),f({\cal  
C}(T))\subseteq{\cal C}(f(T))\bigr\}  
\eeq
that proves to be a quantale $(Q^\#(\Sigma), \bigcup, \circ)$ ---i.e., a join-complete lattice
equipped with an additional operation 
$\circ$ that distributes over arbitrary joins--- where $\bigcup$ is computed pointwise, and which is in
epimorphic ---quantale--- correspondence with the 
`quantale' $(Q({\cal L},{\cal L}), \bigvee, \circ)$ ---see above--- by:
\beq\label{qunatdual}
\bigvee[-]: Q^\#(\Sigma)\to Q({\cal L},{\cal L}):f\mapsto [f_{\cal L}:{\cal C}(T)\mapsto{\cal C}(f(T))]
\eeq
Note that this 
quantale epimorphism indeed exactly expresses that with any state transition there corresponds a join
preserving propagation of properties. We apply all this to the context of this paper: 
\par\smallskip
\par\noindent  
{\it
For each individual entity $S$, let
$\Psi:{\cal L}\to Q^\#(\Sigma)$ be the map that assigns to any property $a\in{\cal L}$ the
---not-necessarily deterministic--- state transition $\Psi(a):{\cal P}(\Sigma)\to {\cal P}(\Sigma)$ that
$S$ undergoes when
actuality of property\, $a$\, is induced on it by interaction with another individual entity.}
\par\smallskip
\par\noindent
Clearly this implicitly determines the
propagation of properties
$\Psi_{{\cal L}}:{\cal L}\to Q({\cal L},{\cal L})$ by $\bigvee[\Psi(-)]=\Psi_{{\cal L}}(-)$. We will now
discuss
$\Psi$ and
$\Psi_{\cal L}$ for ${\cal L}$ orthomodular, and more specific, for the Hilbert space
case.      
\par\bigskip\par  
\noindent 
{\it i) ORDERING OF STATES VIA `DESCENDING' INDUCTION OF ACTUAL PROPERTIES}    
\bpr
\label{orderofstates}  
If the following conditions are satisfied:   
\par\noindent
(i) $\Psi(a)$ does not alter proper states of which the strongest actual property is stronger than
$a$; 
\par\noindent
(ii) all properties compatible to the induced one that are actual beforehand remain actual; 
\par\noindent
(iii)
the assignment\, $im(\Psi_{{\cal L}})\to im(\Psi):\Psi_{{\cal L}}(a)\mapsto\Psi(a)$
preserves composition ---with $im(-)=image$, 
\par\noindent
then,
$p\unlhd q\Leftrightarrow[\exists a\leq{\cal C}(\{q\}):p\in\Psi(a)(\{q\})]$ defines a poset 
$(\Sigma,\unlhd)$ that embeds in
$(\Sigma,\leq_{\cal C})$. 
\epr
\bpf
Following Piron (1976), p.69, Theorem 4.3, if properties described by an orthomodular
property lattice change in such a way that a property $a\in{\cal L}$ becomes actual, and such
that all properties compatible to $a$ that where actual beforehand, are still actual afterwards,
then the corresponding transition of properties is exactly described by the Sasaki projection
$\varphi_{a}:{\cal L}\to{\cal L}:b\mapsto a\wedge(b\vee a^\perp)$. Thus, (ii) assures that $\Psi_{\cal
L}(a)=\varphi_{a}$. If $p\in\Psi(a)(\{q\})$ for $a\leq{\cal C}(\{q\})$, then ${\cal
C}(\{p\})\leq{\cal C}(\Psi(a)(\{q\}))=\varphi_{a}({\cal C}(\{q\})\leq a$ ---the
equality follows from eq.(\ref{qunatdual})--- we have ${\cal C}(\{p\})\leq{\cal C}(\{q\})$
and thus  $p\leq_{\cal C}q$. Note that (i) is equivalent to
$\Psi(a)$ being identical on all
$p\in\Sigma$ such that ${\cal C}(\{p\})\leq a$, and thus yields reflexivity since it forces the
restriction of
$\Psi(a)$ to $\{T\subseteq\Sigma|{\cal C}(T)\leq a\}$ ---which is the only part of the domain
involved in defining $\unlhd$--- to be idempotent. We can't have
$p\lhd q$ and $q\lhd p$ since any transition $\Psi(a)(\{q\})=\{p\}$ requires again by (i) that
${\cal C}(\{p\})\not\leq a$ where ${\cal C}(\{q\})\leq a$. 
Following Foulis (1960), the maps
$\{\varphi_{a}|a\in{\cal L}\}$ are structured in a complete lattice isomorphic to ${\cal L}$ itself when
ordered by
$\varphi_{a'}\leq\varphi_{a}\Leftrightarrow\varphi_{a'}=\varphi_{a'}\varphi_{a}$. Now consider $a\geq
a'$, then 
$\varphi_{a}\geq\varphi_{a'}=\varphi_{a'}\varphi_{a}$, i.e., 
$\varphi_{a'}\varphi_{a}$ only depends on $a'$ and not on $a$.  Thus, (iii) yields transitivity since it
forces $\Psi(a')(\Psi(a)(\{q\}))=\Psi(a')(\{q\})$, and this independent of $a$ provided that $a\geq
a'$.
\epf
Note that condition (ii) says that actual properties are only altered in a minimal way.
Now consider ---under the assumptions of the above proposition--- a chain $a_1\geq a_1'\geq
\ldots$ in
${\cal L}_1$ of consecutive strongest actual properties of $S_1$. By isotonicity of $f_{1,2}$  we have 
$f_{1,2}(a_1)\geq f_{1,2}(a_1')\geq \ldots$ and thus for $\{p_2'\}=\Psi(f_{1,2}(a_1'))(\{p_2\})$
we have $f_{1,2}(a_1)={\cal C}(\{p_2\})\geq{\cal C}(\{p_2'\})\geq\ldots$ for
$p_2$ initial, yielding $p_2\unrhd p_2'\unrhd\ldots$.  As such, the relation
$\unlhd$ expresses evolution of $S_2$ due to 
mutual induction of actuality as
a {\it descending chain of proper states in $\Sigma_2$, with descending strongest actual properties
in ${\cal L}_2$}.   
\par\bigskip\par
\noindent   
{\it ii) PROPER STATES FOR COMPOUND QUANTUM SYSTEMS}              
\par\smallskip\par    
\noindent 
At this point it is  required to propose a candidate for the sets of 
proper states $\Sigma_1$ and $\Sigma_2$ in the
Hilbert space case that allows us to fully recover the description of compound
quantum systems. Let
$P_{a_i}:{\cal H}_i\to{\cal H}_i$ be the orthogonal projector corresponding to
$\varphi_{a_i}$ for $i\in\{1,2\}$ with $a_i$ a closed subspace of ${\cal H}_i$ and set:   
\beq\label{densitystates}
\left\{
\begin{array}{l}
(\Sigma_i, {\cal C}_i)=\Bigl(\{\rho_i:{\cal H}_i\to{\cal H}_i|\rho_i\ is\ a\
density\ operator\}\ ,\ {\cal P}(\Sigma_i)\to{\cal L}_{{\cal
H}_i}:\{\rho_i\}\mapsto\{\rho_i(\phi)|\phi\in{\cal H}_i\}\Bigr)\vspace{1mm}\\
\Psi_i(a_i)=\Bigl[{\cal C}_i(\rho_i)\perp a_i:\{\rho_i\}\mapsto\emptyset\ ;\
{\cal C}_i(\rho_i)\not\perp a_i: \{\rho_i\}\mapsto\{{\scriptstyle{1\over Tr(P_{a_i}\rho_i
P_{a_i})}}P_{a_i}\rho_i P_{a_i}\}\Bigr]   
\end{array}  
\right.
\eeq
One easily verifies that $\Psi_i$ defines state transitions for $(\Sigma_i,{\cal C}_i)$ that fulfill
Proposition
\ref{orderofstates} and that the one dimensional projectors are minimal in
$(\Sigma_i,\unlhd)$. Moreover, the restriction of
$\Psi_i(a_i)$ to $\{T\subseteq\Sigma_i|{\cal C}_i(T)\leq a_i\}$ maps a singleton on a singleton or
$\emptyset$ for all $a_i\in{\cal L}_i$, i.e., the transitions due to induction of properties less
than the strongest actual one are maximally deterministic. Now, given $F\in B({\cal H}_1,{\cal  
H}_2)$ representing
$f_{1,2}$ with
$F^\dagger$ as its adjoint, then:   
\beq\label{FtoQuad}
F\mapsto
\Bigl(F:{\cal H}_1\to{\cal H}_2\ ,\ {\scriptstyle{1\over Tr(F^\dagger F)}}F^\dagger F:{\cal
H}_1\to{\cal H}_1\ ,\ {\scriptstyle{1\over Tr(FF^\dagger)}}FF^\dagger:{\cal H}_1\to{\cal H}_1\ ,\
F^\dagger:{\cal H}_2\to{\cal H}_1\Bigr)
\eeq	
uniquely defines
a quadruple $(f_{1,2},\rho_1,\rho_2,f_{2,1})$ 
which exactly yields the quantum probability structure in the following way:    
(i) The
transition probability for $\{\rho_i\}\mapsto\{\rho_i'\}=\Psi_i(a_i)(\{\rho_i\})$ with
$a_i\not\perp{\cal C}_i(\{\rho_i\})$ in a measurement that verifies the property $a_i$ is
$Tr(P_{a_i}\rho_iP_{a_i})$; (ii) when this happens, say
$\{\rho_1\}\mapsto\{\rho_1'\}$, this transition causes $a_1'={\cal
C}_1(\{\rho_1'\})$ to become actual, and consequently, causes
$a_2'=f_{1,2}(a_1')$ to become actual, having a transition
$\{\rho_2\}\mapsto\{\rho_2'\}=\Psi_2(a_2')(\{\rho_2\})$ as a
consequence; (iii) this reasoning can be proceeded inductively, and
stops once we reach the minimal elements of $(\Sigma_2,\unlhd)$.
It can then be verified that the probability for the chains 
$\rho_1\geq\rho'_1\geq\ldots\geq
P_{\psi}$ and 
$\rho_2\geq\rho'_2\geq\ldots\geq P_{\phi}$ to have a couple $({\cal
C}_1(P_{\psi}),{\cal
C}_2(P_{\phi}))$ as respective outcome `states' ---which are indeed represented as atoms of the
property lattice--- is given by\,:\footnote{A  
proof follows from identification of this construction with the representation for compound
quantum systems in Coecke (1998a): our choice of
$(\Sigma_i,{\cal C}_i)$ and the corresponding quadruples
$(f_{1,2},\rho_1,\rho_2,f_{2,1})$ for linear maps $F$ coincide exactly.}
\beq
{\scriptstyle{1\over ||\psi||_{{\cal H}_1}^2||\phi||_{{\cal H}_2}^2||F||_{HS}^2}}|\langle  
\psi\otimes\phi|\sum_ic_i\psi_i\otimes\phi_i\rangle|^2
\eeq
for $F=\sum_{i=1}^mc_i\langle-|\psi_i\rangle\phi_i$, and this is indeed the quantum transition
probability in a measurement on a compound system described by
$\sum_ic_i\psi_i\otimes\phi_i\in{\cal H}_1\otimes{\cal H}_2$. 
We end by stressing that due to the assignment in eq(\ref{FtoQuad}), the initial proper states
of the compound quantum system are fully encoded in the states of compoundness, and thus encoded in
${\cal H}_1\otimes{\cal H}_2$ via $B'({\cal H}_1,{\cal H}_2)$.     
\par\bigskip\par
\noindent 
{\bf 5. CONCLUSION}
\par\bigskip\par
\noindent
In this paper we proposed an alternative approach towards an understanding of the description
of compound quantum systems, by essentially focusing on the interaction of the individual
entities within the compound system.  Obviously, a lot more investigation could be done on a more
accurate characterization of quantum entanglement as a special case of the primal considerations
made in this paper.  Also an elaboration on the description of compound
systems consisting of more than two entities would be worthwhile.  
\par\bigskip\par
\noindent 
{\bf 6. REFERENCES}     
\par\bigskip\par
\noindent Birkhoff, G. (1940) {\it Lattice Theory}, AMS Coll. Publ.

\noindent Foulis, D.J. (1960) {\it Proc. AMS} {\bf 11}, 648.
 
\noindent Piron, C. (1964) {\it Helv. Phys. Acta} {\bf 37}, 439.

\noindent Pool, J.C.T. (1968) {\it Comm. Math. Phys.} {\bf 9}, 118.
 
\noindent Piron, C. (1976) {\em Foundations of Quantum Physics}, W. A. Benjamin, Inc.   
 
\noindent Hellwig, K.-E. and Krausser D. (1977) {\em Int. J. Theor. Phys.} {\bf 16}, 775.  
 
\noindent Piron, C. (1977) {\em J. Phyl. Logic} {\bf 6}, 481.  
 
\noindent Zecca, A. (1977) {\em J. Math. Phys.} {\bf 19}, 1482.
 
\noindent Aerts, D., and Daubechies, I. (1978) {\it Helv. Phys. Acta} {\bf 51}, 661.
   
\noindent Weidmann, J. (1981) {\em Linear Operators in Hilbert Space}, Springer-Verlag.

\noindent Aerts, D. (1982) {\em Found. Phys.} {\bf 12}, 1131.
 
\noindent Johnstone, P.T. (1982) {\it Stone Spaces}, Cambridge University Press.
 
\noindent Aerts, D. (1984) {\em J. Math. Phys.} {\bf 25}, 1434.
 
\noindent Pulmannov\'a, S. (1984) {\em J. Math. Phys.} {\bf 26}, 1.
 
\noindent Faure, Cl.-A. and Fr\"olicher, A. (1993) {\em Geom. Dedicata} {\bf 47}, 25.
 
\noindent Faure, Cl.-A. and Fr\"orlicher, A. (1994) {\em Geom. Dedicata} {\bf 53}, 273.  
 
\noindent Coecke, B. (1995) {\it Helv. Phys. Acta} {\bf 68}, 394.  
 
\noindent Faure, Cl.-A., Moore, D.J., and Piron, C. (1995) {\em Helv. Phys. Acta} {\bf 68}, 150.  
    
\noindent Moore, D.J. (1995) {\em Helv. Phys. Acta} {\bf 68}, 658.
   
\noindent Coecke, B. (1998a) {\it Found. Phys.} {\bf 28}, 1109\,; quant-ph/0105093.
   
\noindent Coecke, B. (1998b) {\it Found. Phys.} {\bf 28}, 1347\,; quant-ph/0105094.
      
\noindent Coecke, B. and Stubbe, I. (1999a) {\it Found. Phys. Lett} {\bf 12}, 29\,; quant-ph/0008020.  
   
\noindent Coecke, B. and Stubbe, I. (1999b) {\it Int. J. Theor. Phys.} {\bf 38}, 3269.
   
\noindent Moore, D.J. (1999) {\it Stud. Hist. Phil. Mod. Phys.} {\bf 30}, 61.

\noindent Ischi, B. (2000) {\it L'evolution des Syst\`emes S\'epar\'e en Interaction}, 
PhD-thesis, Universit\'e de Gen\`eve.
   
\noindent Coecke, B. and Moore, D.J. (2000)
Operational Galois adjunctions. In:
{\it Current Research in Operational Quantum Logic: Algebras,
Categories and Languages}, B. Coecke, D.J. Moore, A.
Wilce, (Eds.), pp.195--218, Kluwer Academic Publishers\,; quant-ph/0008021.
   
\noindent Valckenborgh, F. (2000) {\it Int J. Theor. Phys.} {\bf 39}, 939.  
   
\end{document}